\documentclass[aip,preprint]{revtex4-1}

\draft 
\usepackage{graphicx}
\usepackage{subcaption}
\usepackage{dcolumn}
\usepackage{bm}

\usepackage[utf8]{inputenc}
\usepackage[T1]{fontenc}
\usepackage{mathptmx}
\usepackage{amsmath}

\DeclareUnicodeCharacter{2212}{-}

\begin{document}

\preprint{}

\title{Multiplex Recurrence Networks from multi-lead ECG data} 



\author{Sneha Kachhara}
\email{snehakachhara@students.iisertirupati.ac.in}
\affiliation{Indian Institute of Science Education and Research (IISER) Tirupati, Tirupati - 517507, India}

\author{G. Ambika}
\email{g.ambika@iisertirupati.ac.in}
\affiliation{Indian Institute of Science Education and Research (IISER) Tirupati, Tirupati - 517507, India}


\date{\today}

\begin{abstract}
We present an integrated approach to analyse the multi-lead ECG data using the frame work of multiplex recurrence networks (MRNs). We explore how their intralayer and interlayer topological features can capture the subtle variations in the recurrence patterns of the underlying spatio-temporal dynamics. We find MRNs from ECG data of healthy cases are significantly more coherent with high mutual information and less divergence between respective degree distributions. In cases of diseases, significant differences in specific measures of similarity between layers are seen. The coherence is affected most in the cases of diseases associated with localized abnormality such as bundle branch block. We note that it is important to do a comprehensive analysis using all the measures to arrive at disease-specific patterns. Our approach is very general and as such can be applied in any other domain where multivariate or multi-channel data are available from highly complex systems.
\end{abstract}
\pacs{05.45.−a, 05.45.Ac, 87.18.−h, 05.90.+m, 05.45.Tp}

\maketitle 

\begin{quotation}
The Electrocardiogram (ECG) is a record of the electrical activity of the heart in the form of a time series. The study of cardiac dynamics through ECG has gathered a lot of attention in the nonlinear dynamics community, with attempts to justify the degree of chaos in this system as well as to identify anomalies in the case of a disease. Nonlinear time series methods, sometimes coupled with Machine Learning approaches, have been employed to this end with reasonable success. However, the interpretation of feature based classification studies in terms of underlying dynamics is not explored, except for some particular ailments such as Arrythmias and Chronic Heart Failure. Moreover, the ECG data comes as highly correlated multivariate data, from 3, 5 or 12 leads. Most of the study from the dynamics point of view till now are on the average data over leads or on a single lead. In the present work we aim to study multi-lead ECG within the framework of Multiplex Recurrence Networks, which highlights spatio-temporal features of the cardiac dynamics as reflected in ECG. To this end, we employ layer similarity/dissimilarity measures in addition to the standard complex network measures defined for multiplex complex networks. We include three levels of structural aspects of MRNs: the coarse structure in terms of links across layers, the interlayer features in degree distributions, and the local micro-structures using local clustering coefficients. We show that the cardiac dynamics in the case of a disease manifests abnormalities in a multitude of ways and can be understood only by consolidated results from a set of measures. 
\end{quotation}
\section{\label{sec:s1}Introduction}

Research related to chaotic behaviour in the cardiovascular system has seen consistent progress in the last two decades\cite{glass2009introduction,cherry2017introduction}.
While the initial attempts were aimed at establishing the presence of deterministic chaos in the cardiac system, techniques of nonlinear time series analysis and machine learning are now being applied to understand cardiac dynamics \cite{acharya2007advances}. The focus has shifted to the identification of signatures of altered dynamics (chaotic or not) in patients as compared to healthy\cite{cherry2017introduction}, leading to good progress in machine learning based diagnostics and early-warning tools. However, this approach relies heavily on the availability of huge databases of quality data to train the algorithms and provides little insight into the underlying dynamics that reflects the intricacies related to cardiac malfunctions.\\

The physiologically relevant signal in the context of the heart is the Electrocardiogram (ECG) which records the electrical activity of the heart\cite{garcia201312}. The data on Heart Rate Variability (HRV) on the other hand, is the number of cardiac cycles per minute. ECG and HRV, both reflect related but different aspects of the cardiovascular dynamics\cite{acharya2007advances,acharya2006heart}. 
While HRV data has the advantage that it is easy to obtain, even with simple wearable devices, the full ECG waveform, on the other hand, is very sensitive to noise, requires specific devices to record and is usually available only for short duration. However, the ECG contains more information about the dynamics, and also at different timescales. As such, analyzing ECG from a dynamical systems' perspective can be quite rewarding and relevant\cite{shekatkar2017detecting}. Studies in this direction indicate reduction of complexity in some diseases such as CHF\cite{banerjee2016complexity} and Arrythmias\cite{qu2011chaos,qu2014nonlinear,tuzcu2006decrease}, however the question of how exactly the dynamics is altered in the case of specific diseases, and measures that reflect these abnormalities from a dynamical point of view, remain relatively unexplored. \\

We note that limitations such as short duration and non-stationarity may have restricted the application of tools of Nonlinear time series analysis to ECG data.
In this context the method of Recurrence Networks (RNs) proves to be useful in analysing short and non-stationary data. It transforms the recurrence pattern in the reconstructed dynamics from a given time series into a complex network\cite{zou2019complex, donner2010recurrence}. In this context we have recently reported bimodality in the degree distribution and scaling of link density with recurrence threshold as characteristic features of RNs from ECG \cite{kachhara2019bimodality}.\\ 

The multivariate data taken from a complex dynamical system has properties that can be uncovered only with a comprehensive approach of analysis. In the case of a multivariate time series, Eroglu et al.\cite{eroglu2018multiplex} have proposed the framework of Multiplex Recurrence Networks in which layers of the network correspond to the different time series of the data. The patterns of connections inside each layer are governed by the dynamics as reflected in the corresponding time series of the original data. The framework is very recent but has been successful with coupled map lattices and regime changes in palaeobotanical data \cite{eroglu2018multiplex}, oil-water spatial flow \cite{gao2017multiplex}, and even for self-reports of human experience (EMA and ESM data)\cite{hasselman2020studying}.\\

In this context, the multi-lead ECG provides multivariate data as it is recorded from the different leads as discrete time series \cite{garcia201312}. There has been some progress in developing diagnostic tools based on 3-lead vectorcardiogram (VCG), which is constructed from the 12-lead ECG and provides a succinct representation of phase relationships. However, in clinical practice 12-lead ECG is preferred\cite{man2015vectorcardiographic} by cardiologists to make diagnosis based on some or all of the leads, depending on the nature of the disease.\\

We start with the hypothesis that the lead-to-lead variations and the similarity underlying data between pairs of leads are to be studied to understand the complexity of the spatio-temporal dynamics of normal heart. Moreover, such a study can give information on any anomaly appearing in a specific set of leads which can be an evidence of a specific disorder. We show how this is feasible by using the framework of multiplex recurrence networks (MRNs). We use the ECG data from the six precordial leads placed closest to the heart as the multivariate data. MRNs can be constructed from multi-lead ECG such that recurrence pattern in the reconstructed dynamics with time series from each lead form one layer of the multiplex, with correspondence between nodes across layers that relate to the same instance in time in the reconstructed dynamics. The analysis of interlayer similarities in such MRNs can uncover patterns not apparent in a single layer alone, as well as provide deeper insight into localized anomalies.\\

We begin by describing briefly the construction of RNs and hence MRNs from the 6 layers of RNs, and the network quantifiers defined on them. We study the MRNs at different levels of complexity; from individual links in layers to the overall topological features. We base the analysis on three quantifiers: links, quantified by average edge overlap; degree distributions compared between pairs of different layers; and distribution of simple cliques, quantified by average local clustering coefficients in different layers. We illustrate how specific diseases can cause differences in measures of similarity between particular pairs of layers, reflecting the intricate variations in the underlying spatio-temporal dynamics.\\

\section{\label{sec:s2}Data and processing}
We use the PTB dignostics\cite{bousseljot1995nutzung,kreiseler1995automatisierte} database from Physionet\cite{goldberger2000physiobank}. In total, data from 125 subjects are used in our analysis, 51 of which are healthy (HC) and the rest of them are from one of the four diseases: Bundle Branch Block (BB, 15), Cardiomyopathy/Heart Failure (CM, 16), Dysrhythmia (DR, 14) and Myocardial Infarction with no secondary diagnosis (MI, 29). Each record originally consists of 12-lead ECG, mostly 60 seconds in duration, sampled at 1000 Hz (60,000 points). We choose data from the precordial leads $v_{1}$ to $v_{6}$ which are placed closest to the heart, for analysis. Each data is pre-processed\cite{berkaya2018survey}; filtered with 0.5-40 Hz, normalized as 0 to 1 and binned to 5000 points. 

\section{\label{sec:s3}Construction of Multiplex Recurrence Networks}

\begin{figure}
    \centering
    \begin{subfigure}{\linewidth}
        \includegraphics[width=0.7\linewidth]{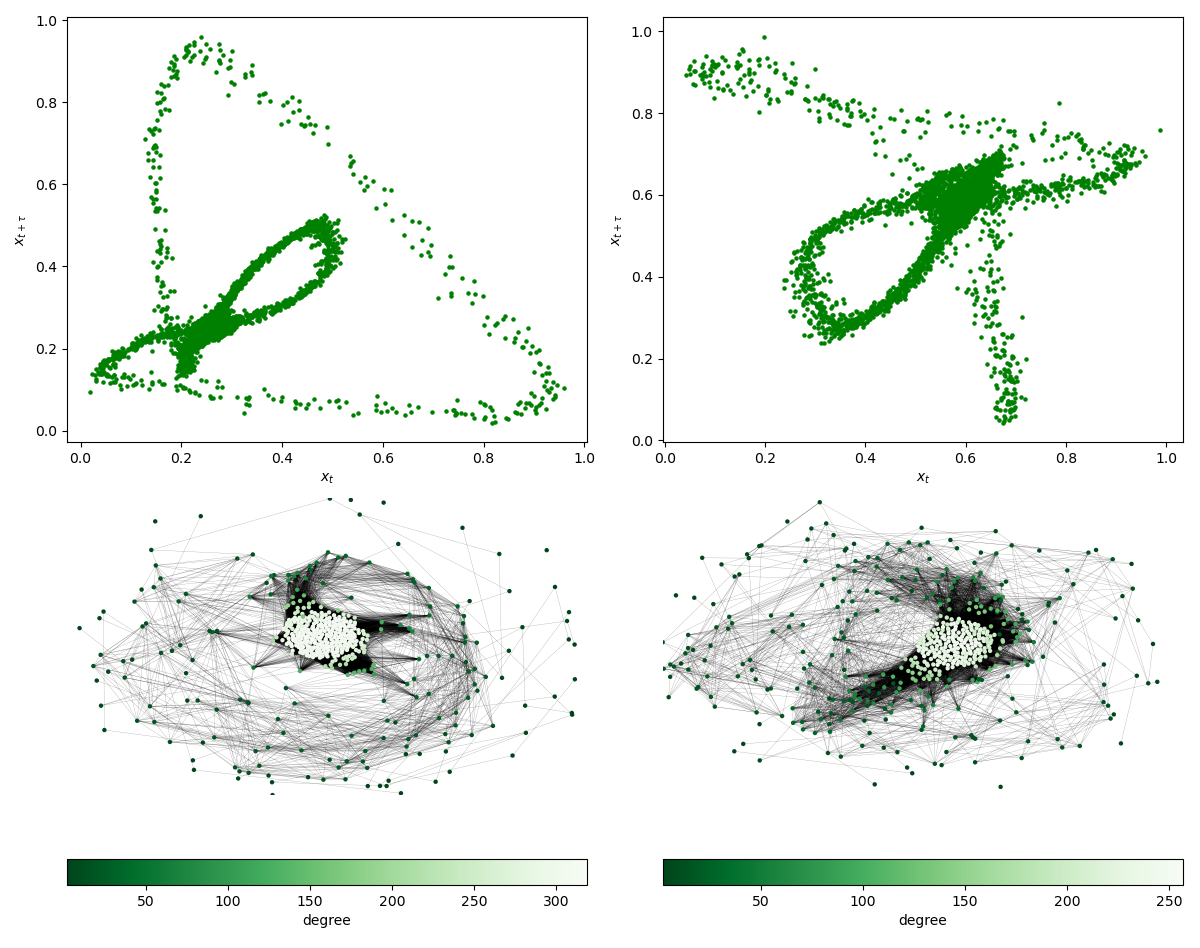}
        \caption{}
    \end{subfigure}
    \begin{subfigure}{\linewidth}
        \includegraphics[width=0.7\linewidth]{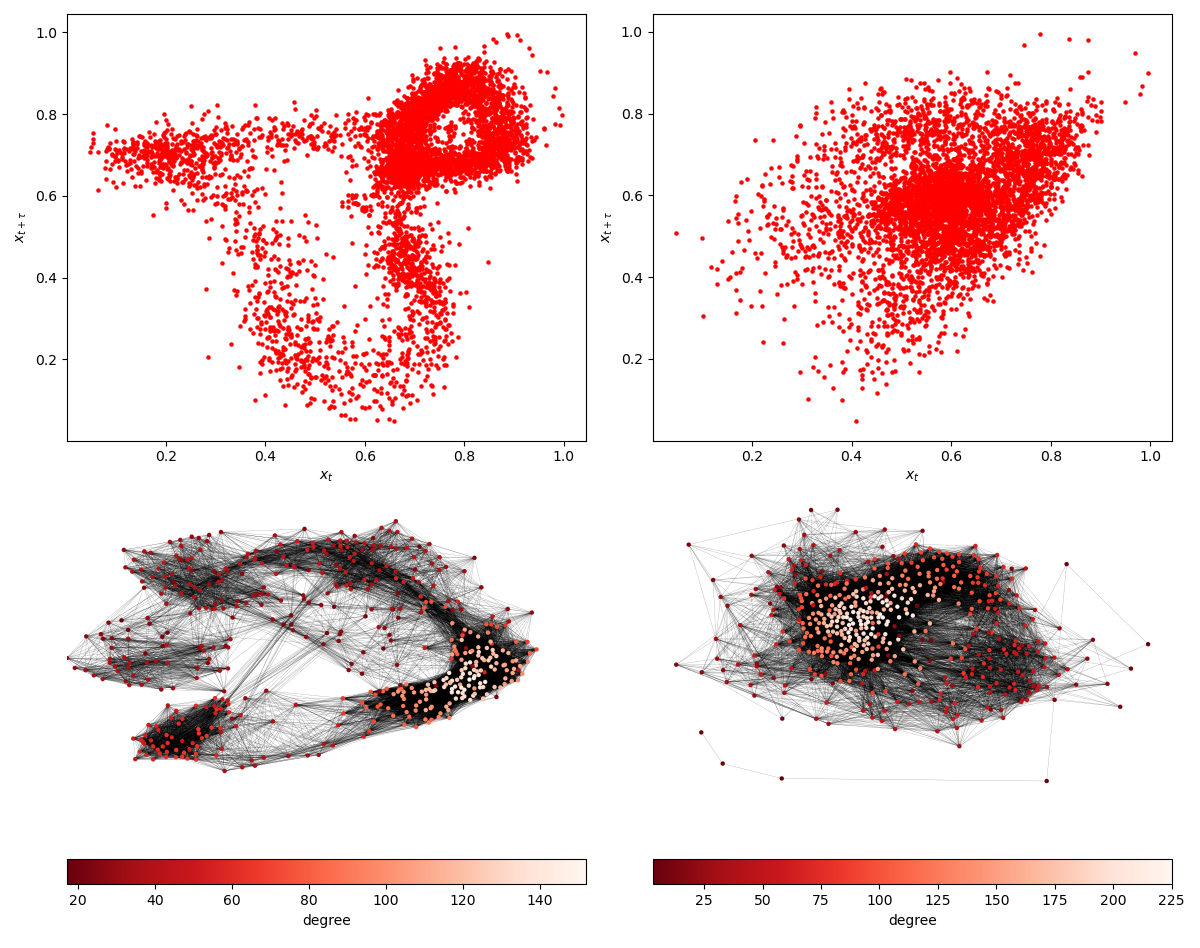}
        \caption{}
    \end{subfigure}
    \caption{Reconstructed attractors in 2-d (top panel) and the corresponding recurrence networks (bottom panel) from two leads 2 (left) and 4 (right) for two typical datasets, (a) healthy and (b) BB. The color of a node indicates its degree.}
    \label{fig:ch24_att_RN}
\end{figure}
The construction of Multiplex Recurrence Networks (MRNs) is done by first constructing the recurrence network from the time series of each lead. This procedure involves two main steps: embedding the time series in a phase space based on  
Taken's embedding theorem\cite{kantz2004nonlinear,ambika2020methods}, and identifying each point on the reconstructed phase space trajectory (or attractor) as node of the network, making connections between them based on their proximity in recurrences \cite{donner2010recurrence}. The reconstruction of the attractor in phase space requires a delay time $\tau$, and embedding dimension $m$ that are to be chosen appropriately for the data under analysis\cite{kraemer2018recurrence,donner2010ambiguities}. For ECG data from each lead, we fix embedding dimension as 4 using the method of False Nearest neighbours (FNN)\cite{hegger1999practical}. The delay time $\tau$ , i.e. the time at which auto correlation falls to 1/e, is chosen as the minimum value of such $\tau$ among the time series from the 6 leads\cite{ambika2020methods}. Then a recurrence threshold ($\varepsilon$) is chosen to construct the links between nodes and to arrive at the adjacency matrix of the recurrence network \cite{schinkel2008selection,jacob2016uniform}. 
The variation of network measures with recurrence threshold itself is an indicator of the complexity of the underlying attractor, as reported recently\cite{kachhara2019bimodality}. In the present work we set it to $\varepsilon$ = 0.1 for the sake of uniformity, based on previous study.\\

For the chosen $\varepsilon$, the recurrence matrix $R$ is constructed such that if two points $i$ and $j$ on the reconstructed attractor lie within distance $\varepsilon$ of each other, we set the corresponding matrix element $R_{ij}$  to be 1, and 0 otherwise. i.e.
\begin{equation}
    \label{eq.1}
    R_{ij} = \Theta \left ( \varepsilon-\left \| \vec{v_{i}}- \vec{v_{j}} \right \| \right )
\end{equation}
where $\vec{v_{i}}$ and $\vec{v_{j}}$ are the corresponding vectors of $i$ and $j$ in the phase space, and $\Theta$ is the Heaviside step function\cite{marwan2007recurrence}.\\

The adjacency matrix $A$ of a Recurrence Network is obtained from $R$ as:
\begin{equation}
    \label{eq.2}
    A_{ij} = R_{ij}-\delta _{ij}
\end{equation}
where $\delta_{ij}$ is the Kronecker delta function that is inserted to avoid self-links in the network. This construction results in an unweighted and undirected network of size N, where N is the number of points on the reconstructed attractor. \\

\begin{figure}
\centering
\begin{subfigure}{\linewidth}
  \includegraphics[width=\linewidth]{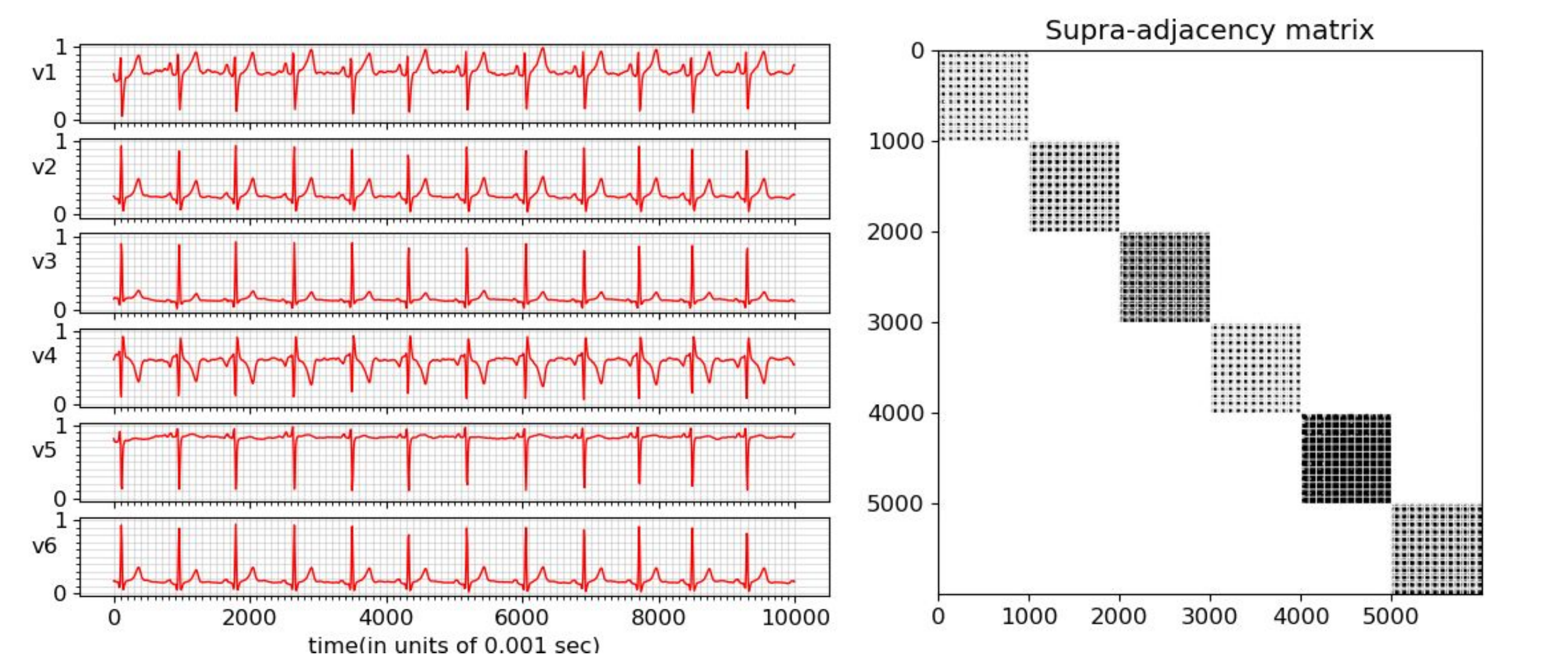}
\end{subfigure}
\begin{subfigure}{\linewidth}
  \includegraphics[width=\linewidth]{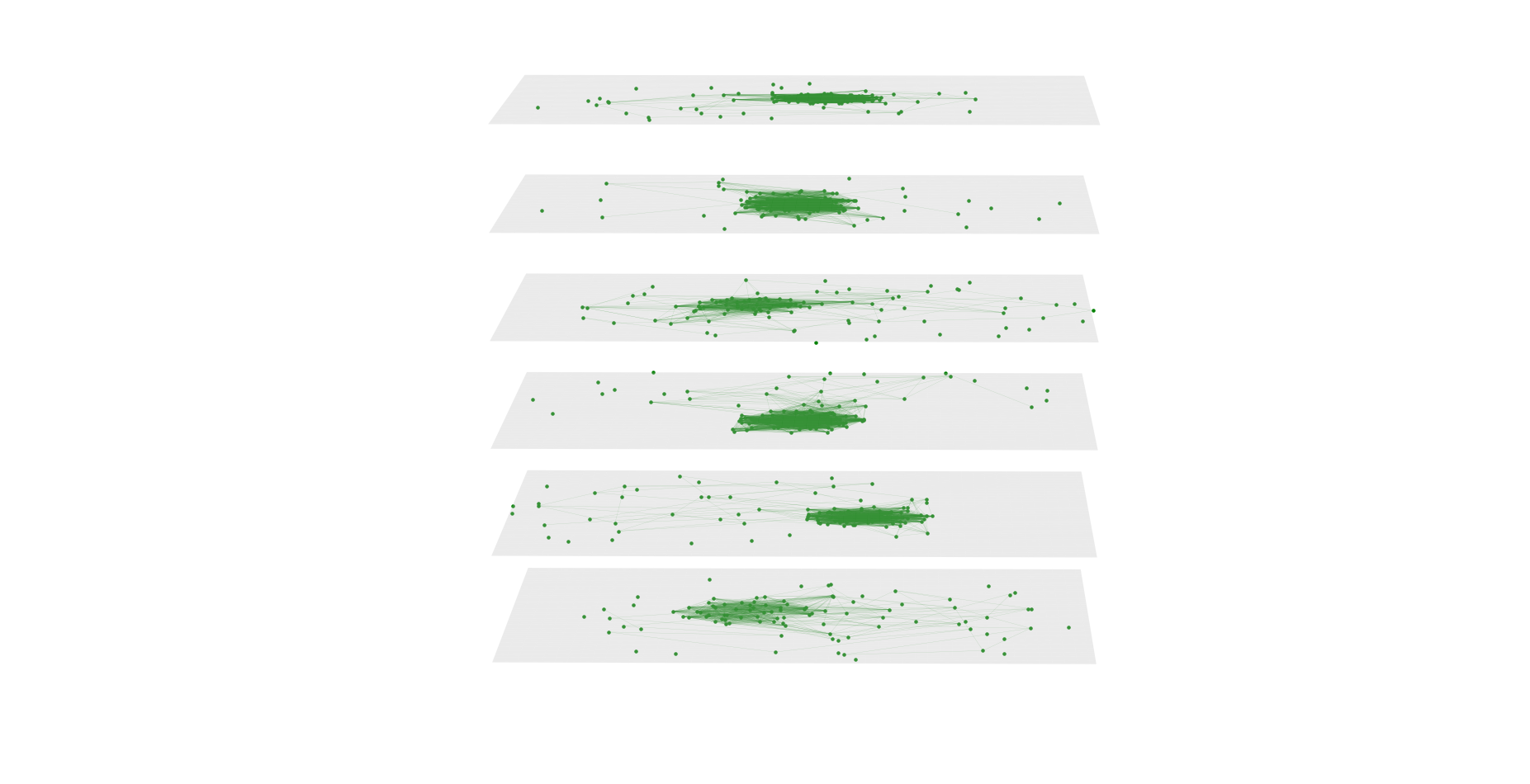}
\end{subfigure}
  \caption{Construction of Multiplex Recurrence Networks from ECG data. The upper panel shows ECG time series and the corresponding supra-adjacency matrix, $M$. The lower panel shows MRN from the matrix. For clarity, the time series and $M$ are shown only for the duration of 10 seconds, and the MRN for 100 nodes. The actual calculations are performed on entire time series, and corresponding MRN has 5000 nodes in each layer.}
  \label{fig:MRN}
\end{figure}
We display in figure \ref{fig:ch24_att_RN} the reconstructed attractors and the corresponding recurrence networks from two different leads for two typical datasets, one of healthy and one of BB. The differences between leads is obvious and more pronounced in the case of a disease. To capture such differences, we construct the MRN by treating RN from each lead as a single layer of a multilayer network,  by connecting the nodes in different layers that correspond to the same instant of time in the reconstructed attractors. 
Thus constructed, the MRN can be represented by a supra-adjacency matrix $M$, which consists of blocks of adjacency matrices from different layers as the diagonal elements and identity matrices as off-diagonal elements:
\begin{equation}
    M = \begin{bmatrix} \mathbf{A}^{[1]} & \mathbf{I}_{N} & \cdots & \mathbf{I_{N}}\\ \mathbf{I_{N}} & \mathbf{A}^{[2]} & \ddots & \vdots \\ \vdots & \ddots & \ddots & \vdots \\ \mathbf{I_{N}} & \cdots & \mathbf{I_{N}} & \mathbf{A}^{[m]} \end{bmatrix}
\end{equation}
Thus it encapsulates properties of the multivariate data effectively, with display of its complex features in the interlayer attributes.
For a typical dataset, we show ECG data from each lead, the supra-adjacency matrix and the constructed MRN in figure \ref{fig:MRN}. We analyze the MRNs with primary focus on interlayer similarities and differences. The quantifiers used in the study are briefly described in the next section and more details can be found elsewhere\cite{brodka2018quantifying, lacasa2015network}.

\section{\label{sec:s4}Measures from Multiplex Recurrence Networks}

We expect a rich structure in multilayer networks because of both interlayer and intralayer connections. For multiplex networks, there is node-to-node correspondence by construction that grants them an additional level of order. Keeping this ordered structure in mind, we select a few measures that can capture their properties. For comparing RNs from two layers, we can consider features at two different levels: local, like degree distribution and global, such as link density (LD)\cite{newman2010networks}. There are other measures such as the average clustering coefficient (CC) which incorporate both local and global structure\cite{newman2010networks}. In addition to these, in the case of multiplex networks, we can effectively use a few other measures defined below which quantify the extent of similarity among the layers. \\

The average edge overlap, denoted by $\omega$, is a consolidated quantifier of the overlap of links (or edges) considering all layers of the network. First proposed by Lacasa et al.\cite{lacasa2015network}, $\omega$ gives an estimate of the expected fraction of layers containing a link, defined as:
\begin{equation}
\label{eq.4.1}
\omega = \frac{\sum_{i}\sum_{j>i}\sum_{l}A_{ij}^{[l]}}{m\sum_{i}\sum_{j>i}(1-\delta _{0,\sum_{l}A_{ij}^{[l]}})}
\end{equation}
where $\delta_{ij}$ stands for the Kronecker delta function, $m$ is the number of layers, and $A_{ij}^{[l]}$ corresponds to the layer $l$ as defined in eq.  (\ref{eq.2}). 
For a network with $m$ layers, $\omega$ by definition, can take values in the range [1/m,1], $\omega$ being 1 only if links are identical in all layers. Thus for MRNs from ECG with 6 layers, $\omega$ will be in [$\frac{1}{6}$, 1].\\

The cosine similarity (CS) is based on the inner product of two vectors and in the present context, in terms of vectors of the degrees of nodes $D^{[l_{1}]}$ and $D^{[l_{2}]}$ in two respective layers $l_{1}$ and $l_{2}$, as: 
\begin{equation}
\label{eq.4.2}
CS = \frac{D^{[l_{1}]}.D^{[l_{2}]}}{{\left \| D^{[l_{1}]} \right \|\left \|D^{[l_{1}]} \right \|}}
\end{equation}
CS is very useful in comparing similarities between layers, node-by-node\cite{brodka2018quantifying}.\\

The index of dissimilarity (ID) calculates differences between two distributions\cite{brodka2018quantifying}. The distributions considered here are that of local clustering coefficients\cite{newman2010networks}. For two layers $l_{1}$ and $l_{2}$, ID is defined as:
\begin{equation}
\label{eq.4.3}
ID = \frac{1}{N} \sum_{i=1}^{N}\left | C_{i}^{[l_{1}]} - C_{i}^{[l_{2}]}\right |
\end{equation}
where $C_{i}^{[l_{1}]}$ and $C_{i}^{[l_{2}]}$ are the local clustering coefficients for the respective layers and N is the total number of nodes.\\

The Jensen-Shannon divergence is a measure to discern two distributions based on the entropy of mixing\cite{pardo2018statistical}. In the context of multiplex networks, we can use the concept of Jensen-Shannon distance ($JSD$) to assess the similarity of the probability degree distributions in different layers. For two layers $l_{1}$ and $l_{2}$, with probability degree distributions $P(k^{l_{1}})$ and $P(k^{l_{2}})$ respectively, we have:
\begin{equation}
\label{eq.4.4}
JSD(P(k^{[l_{1}]})||P(k^{[l_{2}]})) = \frac{\sqrt{D(P(k^{[l_{1}]})||M)+D(P(k^{[l_{2}]})||M)}}{2}
\end{equation}

where $M = \frac{P^{[l_{1}]}+P^{[l_{2}]}}{2}$, is the point-wise mean and $D$ represents Kullback-Leibler divergence\cite{pardo2018statistical}.\\

The interlayer mutual information is calculated on the respective degree distributions as follows\cite{lacasa2015network}:
\begin{equation}
    I_{l_{1}l_{2}} = \sum_{k^{[l_{1}]}}\sum_{k^{[l_{2}]}}P(k^{[l_{1}]},k^{[l_{2}]})\ln \frac{P(k^{[l_{1}]},k^{[l_{2}]})}{P(k^{[l_{1}]})P(k^{[l_{2}]})}
\end{equation}
where $P(k^{[l_{1}]},k^{[l_{2}]})$ is the joint distribution of probability of degrees $k^{[l_{1}]}$ in layer $l_{1}$ and degree $k^{[l_{2}]}$ in layer $l_{2}$. $I_{l_{1}l_{2}}$ captures the topological similarity in the two layers.\\

In addition, we use the Pearson Correlation Coefficient\cite{myers2010research} to compare distributions of local clustering coefficients in two layers. Each of the measures described above except $\omega$ can be averaged over all pairs of layers, which can then be used to compare two MRNs. 
Since the structure of MRN reflects the underlying spatio-temporal dynamics of the cardiac system, any specific variations in measures from two layers and statistically significant changes in a property across subjects of a class will help to understand how diseases can alter heart dynamics and functions. \\

\section{\label{sec:s5}Average measures in MRNs}

\begin{figure}
    \centering
    \includegraphics[width=\linewidth]{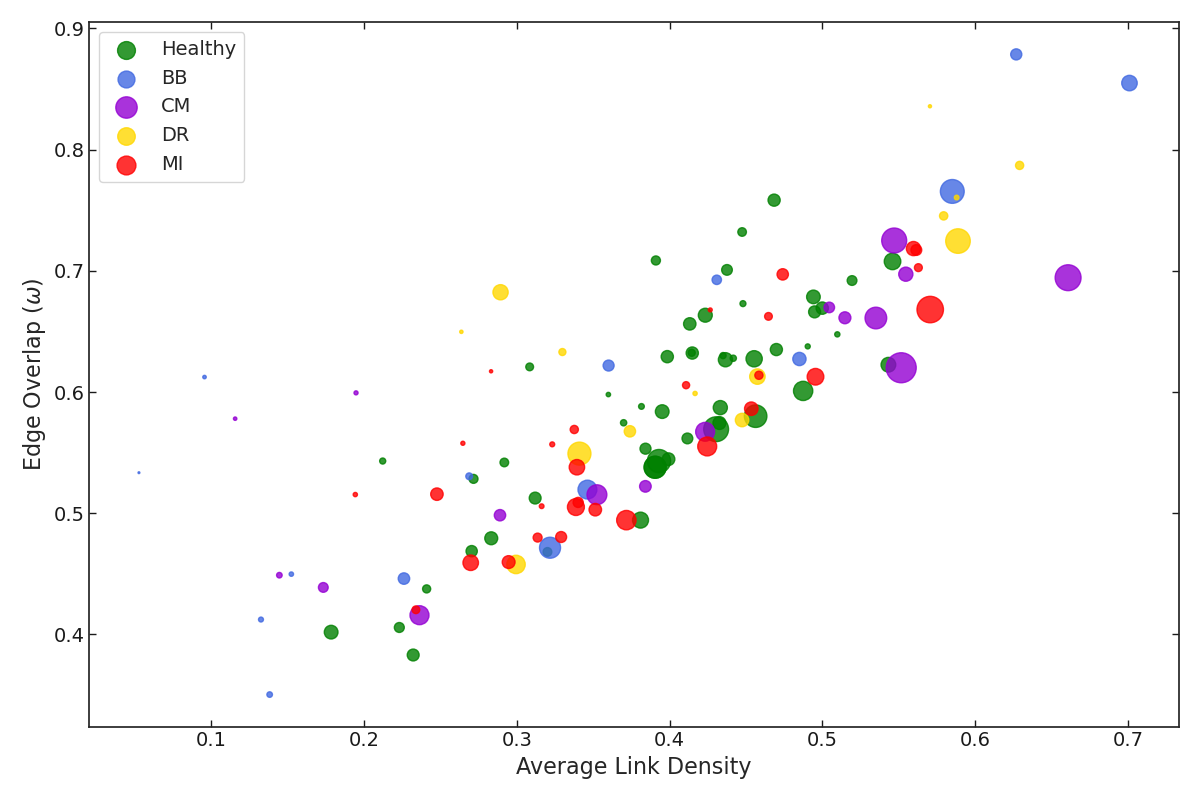}
    \caption{Average edge overlap and average link density across layers of MRNs constructed from ECG data for different subjects. Each circle represents a subject and its size is proportional to the variance in link density across layers. The colors represent different classes of patients, with healthy in green.}
    \label{fig:edge_overlap}
\end{figure}

The one-to-one correspondence of nodes between layers of multiplex networks enables us to compare structures across individual layers on a very basic and concrete level, that of the individual links. In case of MRNs, links across layers can be associated with the recurrences in the underlying dynamics that occur synchronously, making them even more relevant. We compute the average edge overlap, $\omega$ (eq. \ref{eq.4.1}) for the different datasets and the average link density for the 6 layers of MRN from every data set. The results for all the datasets used in the study are shown in figure \ref{fig:edge_overlap}. Each circle represents one subject, and the size of the circle is proportional to the variance in link density (LD) across layers. \\

\begin{table}[htb]
    \centering
    \begin{tabular}{c|c|c|c}
         Category &	edge overlap ($\omega$) &	Average LD	&	Correlation Coefficient	\\
         \hline
         Healthy & 0.5845	$\pm$	0.1533	&	0.3282	$\pm$	0.1967	&	0.8026\\
         BB & 0.5821	$\pm$	0.0966	&	0.3864	$\pm$	0.1692	&	0.8391\\
         CM & 0.6558	$\pm$	0.1015	&	0.4411	$\pm$	0.1245	&	0.7813\\
         DR & 0.5688	$\pm$	0.0842	&	0.3798	$\pm$	0.1048	&	0.6952\\
         MI & 0.5897	$\pm$	0.0852	&	0.3968	$\pm$	0.0881	&	0.8317\\
    \end{tabular}
    \caption{Values of average edge overlap and average link density, for each category of ECG data; along with correlation. The errors indicate standard deviation across subjects of a category.}
    \label{tab:eo_ld}
 \end{table}

We note from figure \ref{fig:edge_overlap} that there is an apparent positive correlation between $\omega$ and average LD. This correlation is further explored using the correlation coefficient for all datasets of a category, as shown in table \ref{tab:eo_ld}. This positive correlation is not surprising since a network with a very high LD naturally has high $\omega$, merely because of increased number of links in all layers. However, their exact relationship depends on how the links are distributed inside the layers. We note that, the correlation is low for DR, and high for MI and BB. 

The results suggest that the MRNs from ECG in general have high degree of association among layers, leading to high values for $\omega$. We note almost all the values for $\omega$ are higher than 0.4, much higher than the lower threshold of $\frac{1}{6} = 0.167$. In some cases, it reaches 0.8-0.9, which is very close to having all layers identical. This would mean that most of the links, or the recurrence points, are common across layers. The healthy cases mostly occupy the middle region, with no healthy subject having higher average LD than 0.6. However, in extreme cases of BB and CM, with either very low or very high average LD, the values of $\omega$ are high (as compared to a healthy person with the same average LD). This leads to the conjecture that there are disproportionate changes in some of the layers in these cases. We will explore this possibility further with the interlayer similarity measures in section \ref{sec:s6}.\\

\section{\label{sec:s6}Inter-layer similarities of degrees}
In this section we discuss the measures to relate the degree distributions across layers. We first discuss how degree of each node differs from layer to layer as captured by the Cosine Similarity or $CS$, and then, how the overall distribution of degrees differs from layer to layer through Jensen-Shannon Distance, $JSD$ and Mutual Information, $I$.\\

\begin{figure}
    \centering
    \includegraphics[width=\linewidth]{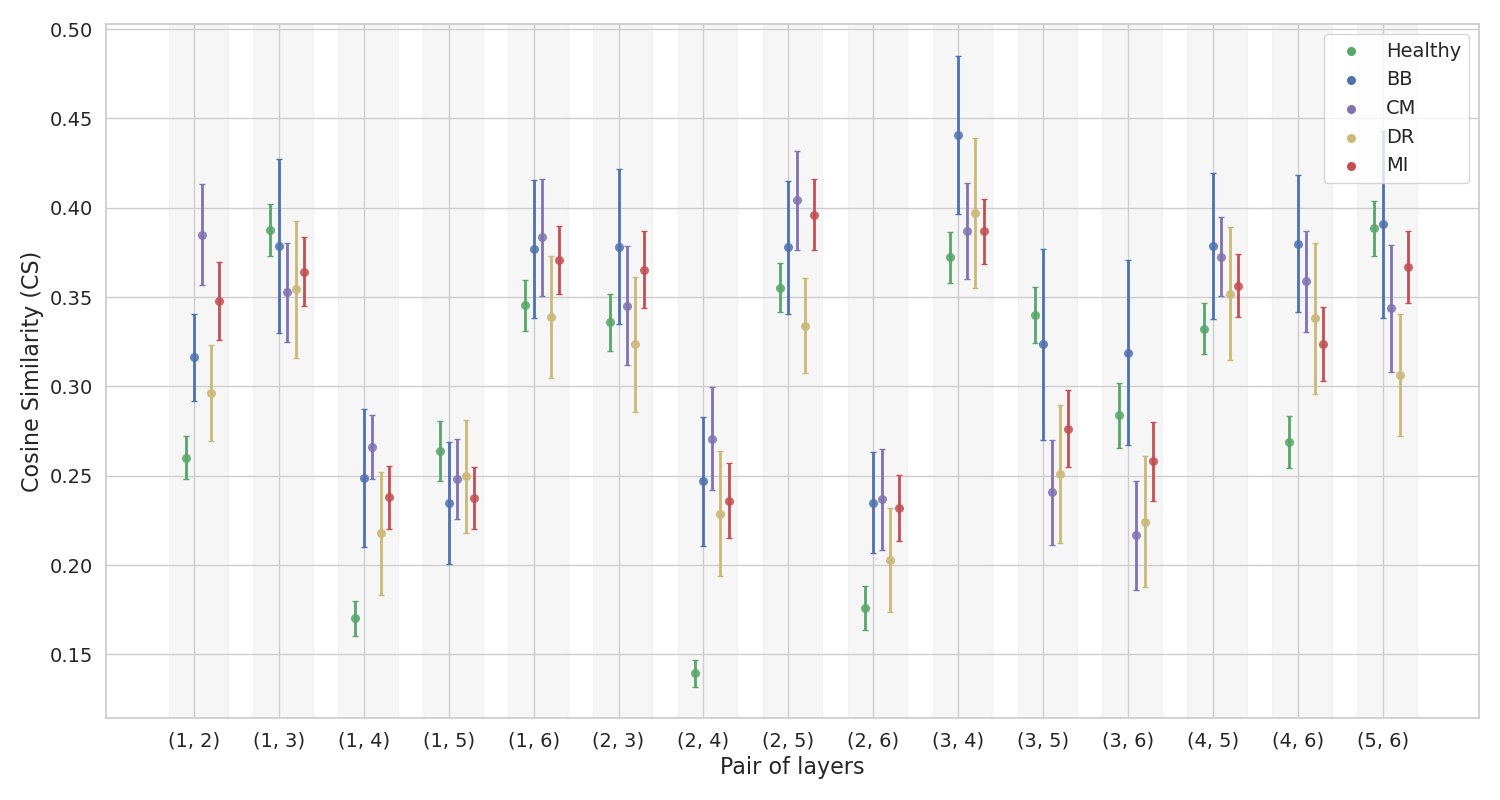}
    \caption{Cosine Similarity of degrees among pairs of layers of MRNs constructed from ECG data. Each point represents the average value of $CS$ for the corresponding pair of layers, across subjects of a category. The colors indicate different categories as before with healthy in green. The errorbars indicate standard error.}
    \label{fig:CS_deg}
\end{figure}

\begin{figure}
    \begin{subfigure}{\linewidth}
     \includegraphics[width=\linewidth]{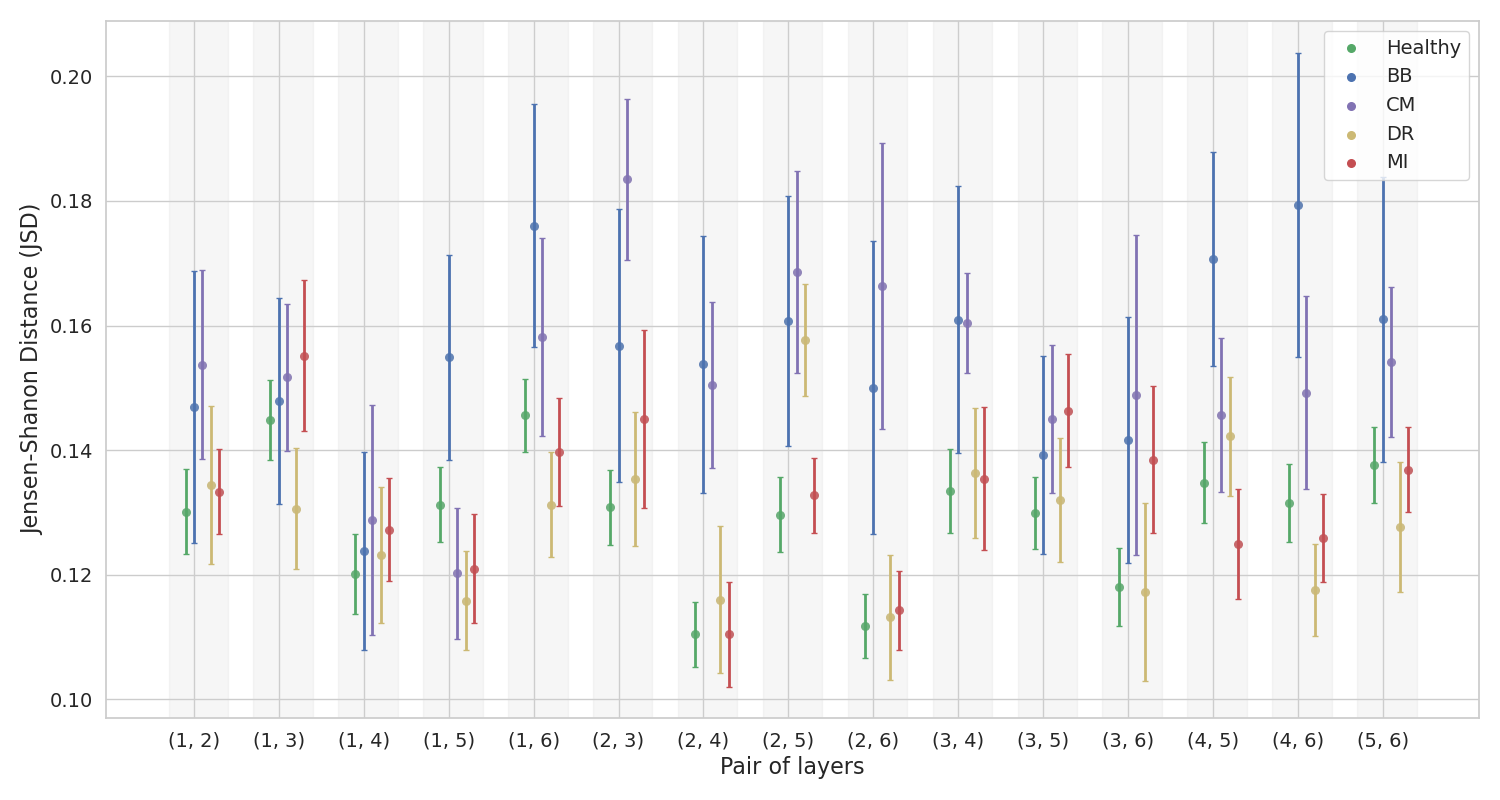}
    \caption{}
    \end{subfigure}
    \begin{subfigure}{\linewidth}
     \includegraphics[width=\linewidth]{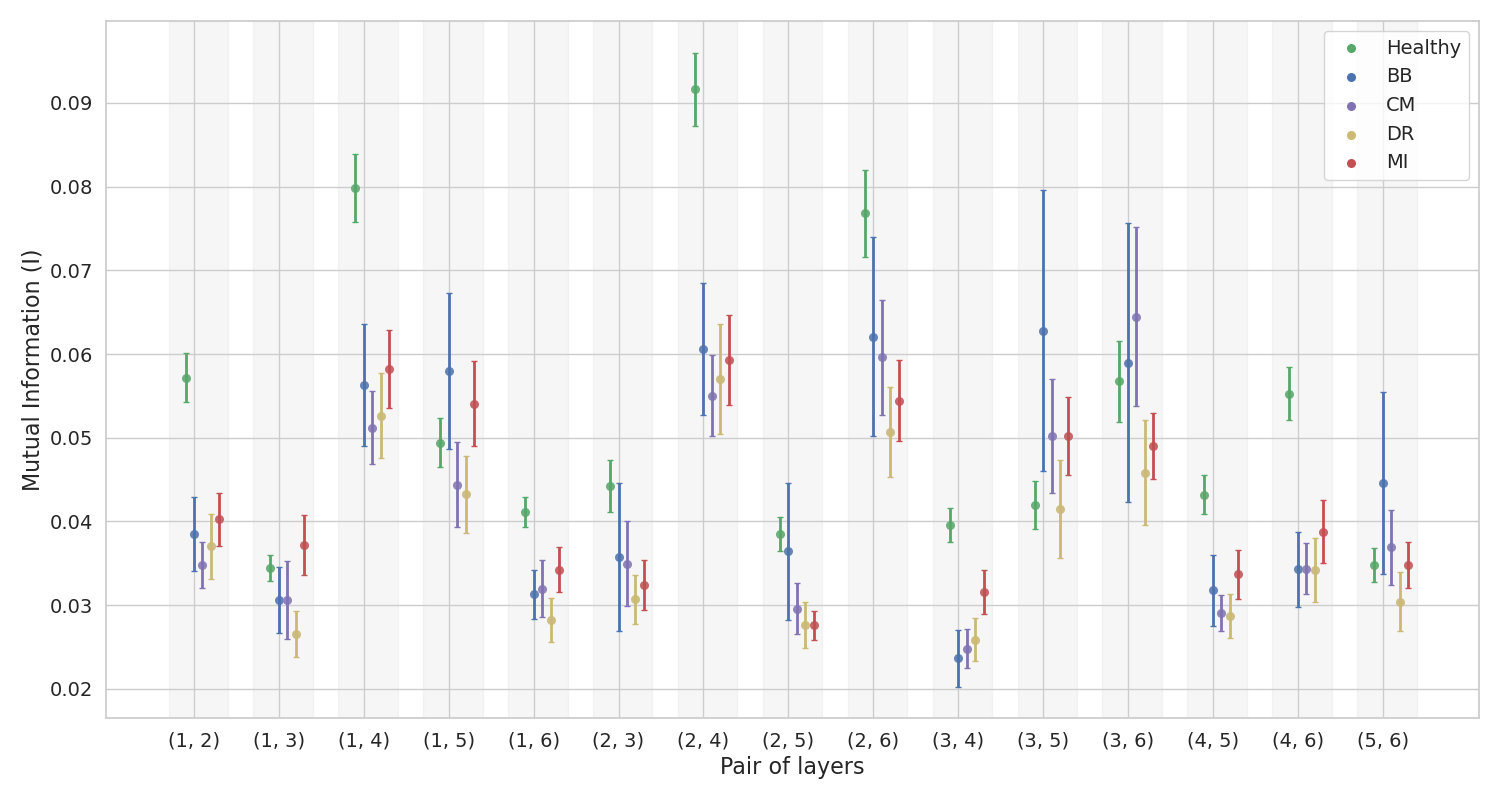}
    \caption{}
    \end{subfigure}
    \caption{(a) Jensen-Shannon distance and (b) Mutual Information from degree distributions among pairs of layers of MRNs constructed from ECG data. Each point represents the average value for the corresponding pair of layers, across subjects of a category. The respective colors correspond to different categories as before, with healthy in green. The error bars indicate standard error.}
    \label{fig:deg_dist_measures}
\end{figure}

The $CS$ by definition, is a local measure and captures node-to-node differences across layers. We compute the $CS$ values taking leads pairwise as (1,2), (1,3) etc. for each dataset. We present the average values of $CS$ computed pair wise for data of each category in figure \ref{fig:CS_deg}, along with standard error. We find that for layer pairs (1,2), (1,4), (2,4), (2,6) and (4,6), the values from healthy are the lowest and in general BB has the highest value in all layers.

\begin{table}[htb]
    \centering
    \begin{tabular}{c|c|c|c}
         Category &	CS &	$JSD$	&	$I$\\
         \hline
         Healthy	&	0.1626	$\pm$	0.0189	&	0.1293	$\pm$	0.0104	&	0.0523	$\pm$	0.0176\\
         BB	&	0.1844	$\pm$	0.0063	&	0.1549	$\pm$	0.0145	&	0.0444	$\pm$	0.0138\\
         CM	&	0.1480	$\pm$	0.0099	&	0.1523	$\pm$	0.0152	&	0.0408	$\pm$	0.0124\\
         DR	&	0.1469	$\pm$	0.0060	&	0.1287	$\pm$	0.0121	&	0.0374	$\pm$	0.0104\\
         MI	&	0.1664	$\pm$	0.0074	&	0.1324	$\pm$	0.0121	&	0.0424	$\pm$	0.0107\\
    \end{tabular}
    \caption{Interlayer similarity measures $CS$, $JSD$ and $I$ in degree distributions, averaged across every pair of layers within each category of data sets. The errors indicate standard deviation across subjects of the category.}
    \label{tab:deg_measures}
\end{table}
 Similarly the values for $JSD$ and $I$ are calculated pairwise and presented in figure \ref{fig:deg_dist_measures} (a) and (b) respectively. 
The values of $I$ for pairs of layers (1,2), (1,4), (2,4), (2,6) and (4,6) are highest for healthy.

All the three measures are averaged over all pairs in each data set and then their means and standard deviations for all datasets in each category are tabulated in table \ref{tab:deg_measures}. We see that $JSD$ is higher for BB and CM, while for DR and MI it stays close to the corresponding range for healthy. As for $I$, healthy have a high value for most of the pairs of layers. \\

\section{Interlayer similarity as reflected in local CC}
\begin{figure}
    \centering
    \includegraphics[width=\linewidth]{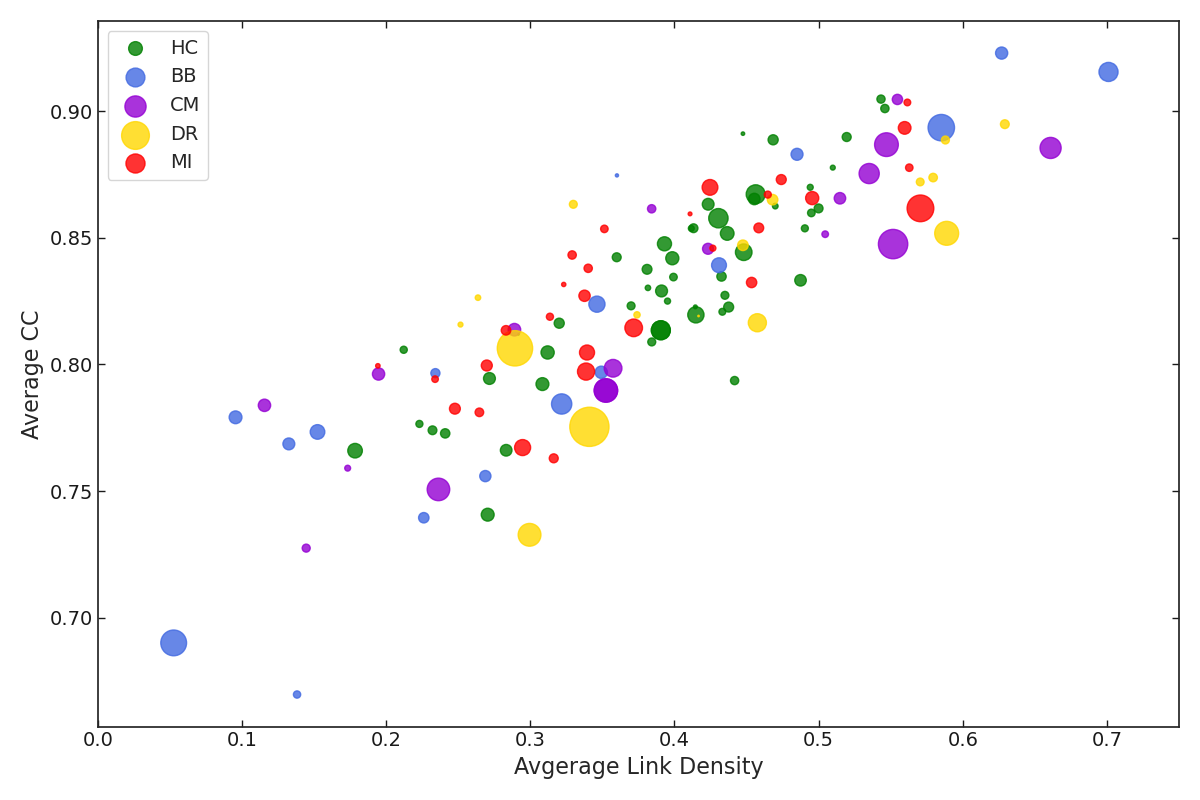}
    \caption{Values of average Clustering Coefficient (CC) and average link density of different subjects. Each circle represents a subject and its size indicates variance in CC across layers for that subject. The colors are representative of the category the subject belongs to, with healthy in green.}
    \label{fig:CC_vs_LD}
\end{figure}

\begin{figure}
    \begin{subfigure}{\linewidth}
    \includegraphics[width=\linewidth]{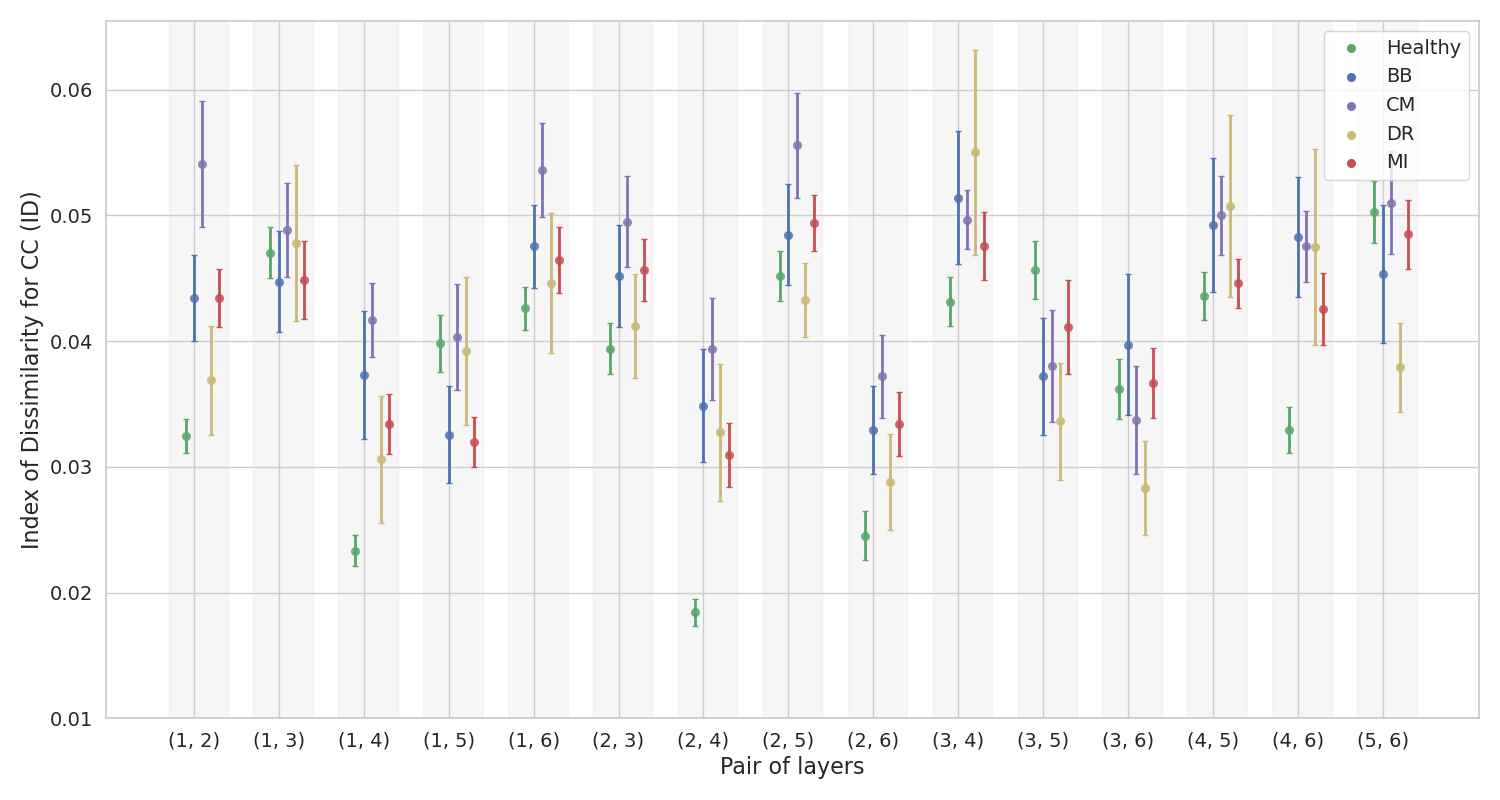}
    \caption{}
    \end{subfigure}
    \begin{subfigure}{\linewidth}
    \includegraphics[width=\linewidth]{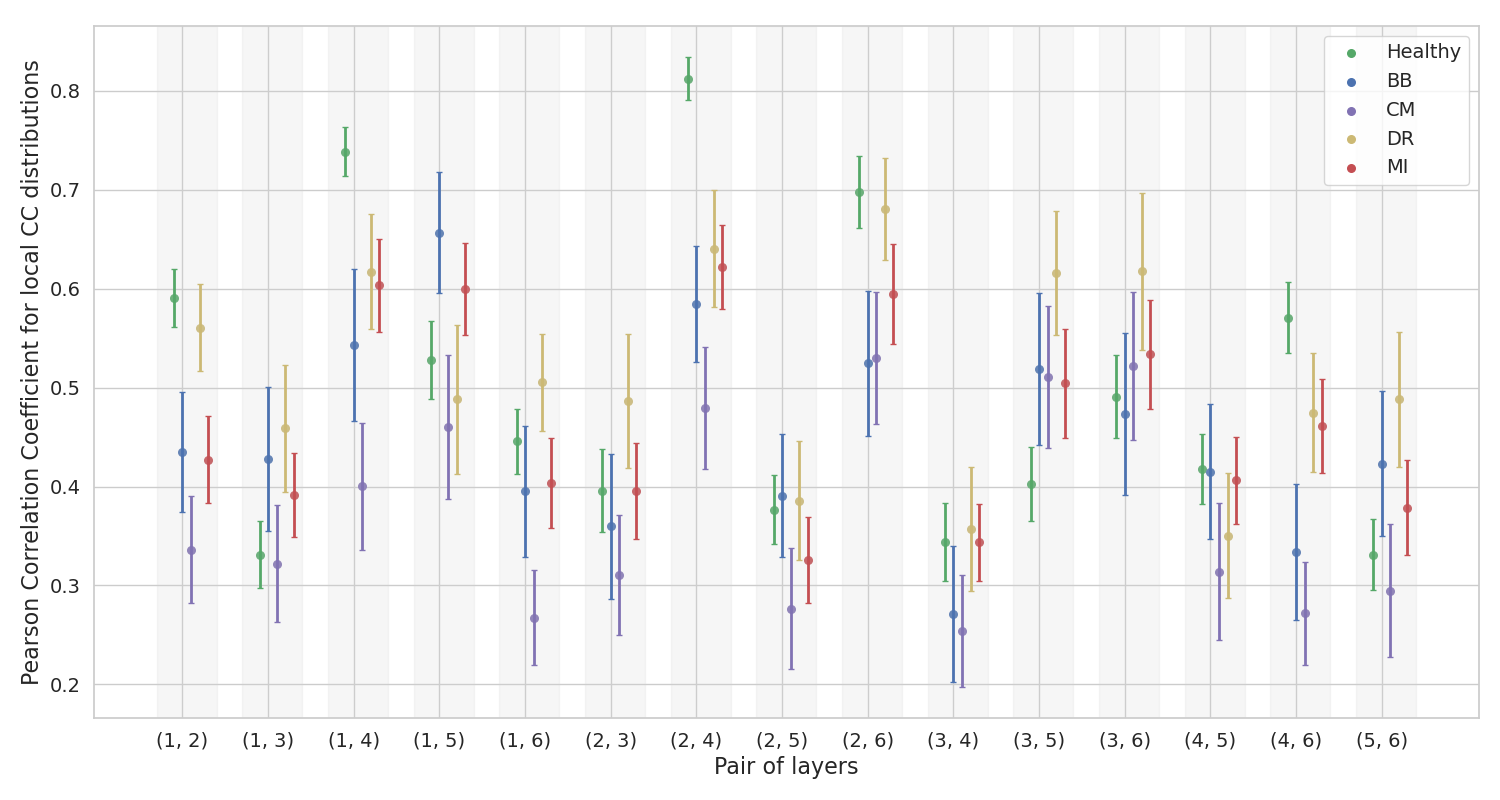}
    \caption{}
    \end{subfigure}
    \caption{(a) Index of Dissimilarity ($ID$) and (b) Pearson Correlation Coefficient for local CC distributions among pair of layers of MRNs constructed from ECG data. Each point in the plot represents average value of the corresponding measure for the pair of layers across different subjects of a category (color-coded by category).}
    \label{fig:CC_measures}
\end{figure}

In this section, we compute the average value of local Clustering Coefficients (CC) of each layer of the MRN in order to analyze the micro-structure. We take the average CC for all the layers and plot them against average LD for every dataset, as shown in the figure \ref{fig:CC_vs_LD}. Each circle in the figure represents a subject and its size is proportional to the variance in average CC across layers for that subject. We note that the magnitude of variance is low for healthy subjects. Also, the average CC and LD are correlated, as reflected in the correlation coefficient summarized in table \ref{tab:cc_measures}. Specifically, for DR, we observe that the values lie significantly away from the main diagonal and hence differ from healthy and MI.\\

\begin{table}[htb]
    \centering
    \begin{tabular}{c|c|c}
         Category 	&	$ID$ & Correlation Coefficient\\
         \hline
         Healthy		&	0.0395	$\pm$	0.0096	&	0.4982	$\pm$	0.1579	\\
         BB		&	0.0437	$\pm$	0.0080	&	0.4445	$\pm$	0.1809	\\
         CM	&	0.0484	$\pm$	0.0065	&	0.3847	$\pm$	0.1554	\\
         DR	&		0.0451	$\pm$	0.0144	&	0.5077	$\pm$	0.1602	\\
         MI	&		0.0414	$\pm$	0.0082	&	0.4661	$\pm$	0.1540	\\

    \end{tabular}
    \caption{Interlayer similarity measures in local CC distributions, averaged across every pair of layers of MRNs for each category. The errors indicate standard deviation across subjects of a category.}
    \label{tab:cc_measures}
\end{table}

Now to depict the dissimilarity in interlayer topology, we compute the Index of Dissimilarity ($ID$), which is a measure of the differences in the two distributions of local CC values in two layers. The results are shown in figure \ref{fig:CC_measures} (a). We can also compare how correlated the distributions are, as depicted using Pearson Correlation Coefficients in figure \ref{fig:CC_measures} (b). The average value for each category over all datasets for all pairs of layers is presented in table \ref{tab:cc_measures}. We find that in general, ID is higher than healthy for all diseases, but most prominently so for CM. Correspondingly, the correlation is also low for CM. \\

\section{Variation in measures among pairs of layers}
\begin{figure}
 
  \begin{subfigure}{0.5\linewidth}
    \includegraphics[width=\linewidth]{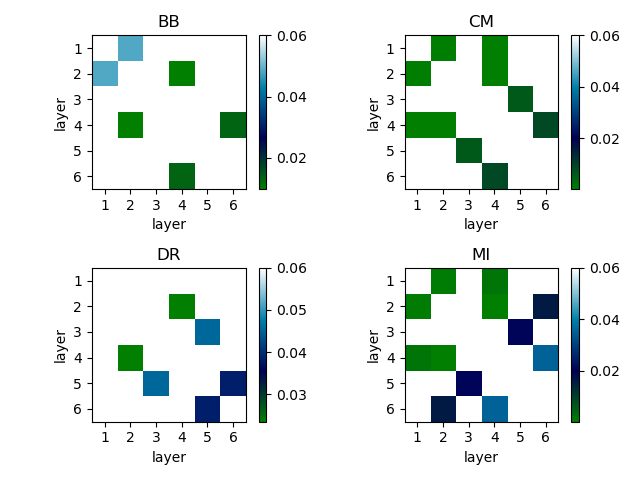}
    \caption{CS (degrees)}
  \end{subfigure}
  \begin{subfigure}{0.5\linewidth}
    \includegraphics[width=\linewidth]{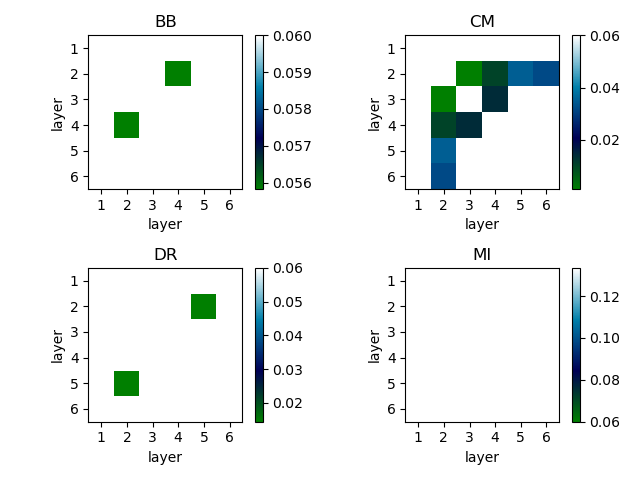}
    \caption{JSD (degree distribution)}
  \end{subfigure}\hfill
  \begin{subfigure}{0.5\linewidth}
    \includegraphics[width=\linewidth]{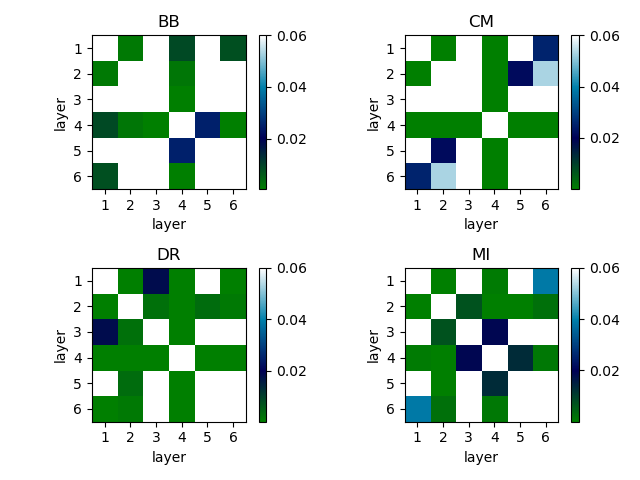}
    \caption{I (degree distribution)}
  \end{subfigure}
  \begin{subfigure}{0.5\linewidth}
    \includegraphics[width=\linewidth]{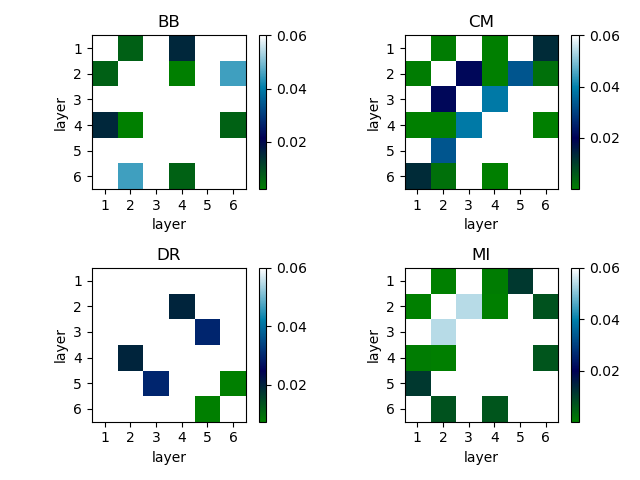}
    \caption{ID (CC)}
  \end{subfigure}\hfill
\caption{Significance of computed measures for different pair of layers of the MRNs of each disease, as compared to healthy. Each off-diagonal matrix element corresponds to a particular pair of layers as indicated by its row and column number. The color code is as per the p-value computed from Welch's t-test. Statistically most significant differences are shown in green (less so in blue), and most insignificant in white.}
\label{fig:sig_res}
\end{figure}

The different measures computed for MRNs from multi-lead ECG data and presented in the above sections are further analysed statistically and consolidated together in this section. For this we use the Welch's t- test\cite{myers2010research} to compute a significance value for every measure for each category and pair of layers, so that we can understand how significant the differences in computed measures are for each disease from healthy. The results are summarized in the form of significance matrices in figure \ref{fig:sig_res}. Each entry in a given matrix represents the significance value for that category as compared to the corresponding measure for healthy, for the pair of layers indicated. A p-value < 0.05 is color coded (p > 0.05 is white) from blue to green in decreasing order such that green indicates the least p-value, corresponding to the most significance.\\

From the figure, we can conclude that different classes of diseases have different types of variations in the cardiac dynamics and not all measures show differences for all diseases. For BB we observe isolated elements in the matrix of significance for measures $CS$, $JSD$ and $ID$. For CM, there seems to be a gradual change across adjacent layers, reflected strongly in $JSD$. For DR, there are few elements in green in all measures except in $I$, and for MI, significant difference is seen for multiple elements in both $I$ and $ID$. No pair of layers show any significant difference in $JSD$ for MI, hence the corresponding matrix is completely white. These results indicate that a disease like BB could be localized, hence some layers are affected more than others, while DR and MI manifest in an integrated manner affecting all the leads. Since each measure encapsulates similarity of a different kind, we expect them to differ from disease to disease.\\

\section{conclusions}
The spatio-temporal features of the electrical excitation patterns of heart are complex and the multiplex recurrence networks is the ideal framework for understanding variations in their complexity. In this study, we analyse the multi-lead ECG data from 125 subjects by constructing multiplex recurrence networks (MRNs) in order to discern patterns of variations in the underlying cardiac dynamics. This includes 51 healthy subjects, and disease cases like Bundle Branch Block (BB), Cardiomyopathy (CM), Dysrhythmia (DR) and Myocardial Infarction (MI). \\

The measures specific to multiplex networks, such as Edge overlap ($\omega$) and Mutual Information ($I$) are highly relevant for ECG data as they can highlight features of cardiac dynamics obscured by inherent non-linearity and correlations in data. Measures such as Cosine Similarity ($CS$) and Jensen-Shannon Distance ($JSD$) can provide additional insight into the topological differences across layers. Moreover, the differences in distribution of local clustering coefficients (CC), captured by the Index of Dissimilarity ($ID$) and correlation coefficient, can provide more information on the interlayer similarities in their micro-structures.\\

The results on $\omega$ and average link density establish that MRNs from ECG have high similarity from layer-to-layer at the most basic level. $\omega$ for healthy is found between 0.4-0.8, and is generally higher as compared to patients for the same average link density. Moreover, most cases of patients with high $\omega$, show large variation in link density from layer to layer. Our results illustrate that healthy cardiac dynamics has less range of variations across layers, as compared to diseases. Most extreme cases are those of BB and CM, while DR and MI are mostly within the range for healthy. The extreme values observed could be either due to some of the layers being very similar, or vice-a-versa.  The value of $CS$, which measure variations in degrees of nodes across layers, has the highest value for BB, as compared to other diseases and healthy. \\ 

We extend the study to interlayer similarities as reflected in degree distributions and distributions of local CCs. We find that the $JSD$ is consistently high for BB and CM among all pairs of layers, while for DR and MI the values are closer to healthy. On the other hand, $I$ is highest for healthy, on the average across all pairs of layers. In particular, we observe that the layer 4 differs most from other layers in the case of BB and CM. For DR and MI, no particular layer or pair of layers stand out in terms of differences in degree distributions. We note that the healthy do not differ much from layer to layer in terms of average local CCs. Thus, we can infer that there is an overall coherence in the healthy cardiac system which is hampered in case of a disease. There are some specific differences in the way different abnormalities manifest in cardiac dynamics. A localized anomaly, such as that of BB for example, will affect only some of the layers but in case of MI, all the layers show significant differences. \\

Our study is aimed at exploring the nature of variations in the underlying spatio-temporal dynamics due to any type of disease as revealed through the measures computed from the multi-lead ECG data. Such an understanding of the dynamics is basic to a proper knowledge of cardiac system and can provide insight for intelligent algorithms. Further studies in this direction can combine deep learning approaches with dynamical systems theory. Moreover, the type of analysis presented here can be applied to multivariate data in other domains where traditional approaches of data analysis are insufficient. The framework of MRNs is not limited to time series data, as the measures can be employed to analyze similarities in any real-world multiplex networks and it will be interesting to see applications in more real data-based networks.

\section{Data Availability}
The data that support the findings of this study are openly available in PTB diagnostics ECG database at https://doi.org/10.13026/C28C71, ref.\cite{bousseljot1995nutzung,kreiseler1995automatisierte}.

\begin{acknowledgments}
One of the authors, Sneha Kachhara, acknowledges financial support from the Council of Scientific and Industrial Research (CSIR), India through Senior Research Fellowship.
\end{acknowledgments}

%
\end{document}